\documentclass[12pt,preprint]{aastex}
\usepackage{natbib}
\usepackage{longtable}
\begin{document}
\title{Disk reflection signatures in the
  spectrum  of  the  bright   Z-source  GX  340+0
    \footnote{Based  on
    observations obtained with XMM-Newton, an ESA science mission with
    instruments and contributions directly funded by ESA Member States
    and NASA}}

\author{A. D'A\`i \altaffilmark{1}, R. Iaria\altaffilmark{1}, 
  T. Di Salvo\altaffilmark{1}, G. Matt \altaffilmark{2}
  N. R. Robba\altaffilmark{1}} 

\altaffiltext{1}{Dipartimento di Scienze Fisiche ed Astronomiche,
  Universit\`a di Palermo, via Archirafi 36 - 90123 Palermo, Italy; 
  email:dai@fisica.unipa.it}
\altaffiltext{2}{Dipartimento di Fisica, Universit\`a degli Studi Roma Tre, 
via della Vasca Navale 84, I-00146 Roma, Italy}
 
\begin{abstract}
  
  We  present the  preliminary  results  of a  50  ks long  XMM-Newton
  observation of the bright Z-source GX 340+0. In this Letter we focus
  on the study of a broad asymmetric emission line in the Fe K$\alpha$
  energy band, whose  shape is clearly resolved and  compatible with a
  relativistically smeared  profile arising  from reflection on  a hot
  accretion  disk extending  close  to the  central accreting  neutron
  star. By  combining temporal and  spectral analysis, we are  able to
  follow  the evolution  of the  source along  its  Horizontal Branch.
  However, despite a significant  change in the continuum emission and
  luminosity,  the line profile  does not  show any  strong correlated
  variation.  This  broad line is produced by  recombination of highly
  ionized iron (Fe XXV) at an inferred inner radius close to 13 R$_g$,
  while the fit requires a high  value for the outer disk radius.  The
  inclination of the  source is extremely well constrained  at 35 deg,
  while the emissivity index is -2.50.
  
\end{abstract}
\maketitle

\section{Introduction}

GX 340+0  is a  bright Low-Mass X-ray  Binary (LMXB) belonging  to the
class of the Z sources \citep{hasinger89}, and its inferred luminosity
is close to the Eddington limit  (2 $\times$ 10$^{38}$ erg/s for a 1.4
M$_{\odot}$ NS).  The source has  also a radio counterpart, from which
a distance  of 11  $\pm$ 3 kpc  has been  estimated \citep{penninx93};
radio emission from the  counterpart is, however, highly variable, and
seems to be  correlated with the X-ray flux when the  source is in its
Horizontal  Branch  (HB), but  anticorrelated  on  the other  branches
\citep{oosterbroek94}; it  can also  show periods of  radio quenching,
becoming extremely faint \citep{berendsen00}.

Temporal analysis  studies have shown a  complex phenomenology, linked
to the  accretion state, with  characteristics typical of  the Z-class
\citep{jonker00}. The  power spectrum  shows a low-frequency  (tens of
Hz)  quasi-periodic oscillation  when  the source  resides  on its  HB
(called   Horizontal  Branch   Oscillation,  HBO)   while   at  higher
frequencies  the   source  shows,  at  the  same   time,  twin  kiloHz
Quasi-Periodic Oscillations (kHz QPOs), whose centroid frequencies are
correlated with  the HBO peak  frequency.  Although no burst  has been
yet observed, the  very fast timing variability (at  more than 800 Hz)
and  the  similarities with  the  other  Z-sources  indicate that  the
compact object is a neutron star.

The spectral properties of the source have not been fully investigated
so far;  \citet{schulz93} studied the 2--12 keV  spectrum using EXOSAT
data; the spectrum  could be well described by  a single component due
to  thermal  Comptonization of  soft  photons,  emerging  from the  NS
surface,  in  a hot  corona  of  moderately  optical thickness  ($\tau
\simeq$ 5--6);

\citet{lavagetto04}  presented  the  first  broadband  (0.1--200  keV)
spectrum of  the source  using data from  a BeppoSAX  observation. The
spectrum could be decomposed into  the sum of a soft thermal component
of temperature  of 0.5 keV, an optically  thick Comptonized component,
and an excess at energies above  20 keV that they fitted with a simple
power law.   A high resolution spectrum  of the source  was studied by
\citet{ueda05},  using  a Chandra  observation;  Chandra data  clearly
showed the presence of an emission line, fitted with a simple Gaussian
profile, at 6.57 keV with 40 eV equivalent width.

In this Letter  we report the results of  an XMM-Newton observation of
GX  340+0.   Thanks  to  the  large collecting  area  of  the  EPIC-PN
instrument and  its good  spectral resolution, we  find evidence  of a
strong broad emission line at $\sim$ 6.7 keV; the line profile is very
well reproduce by a smeared disk-reflected profile; we find, moreover,
evidence  ot other  disk reflection  signatures supporting  the common
scenario, valid  for AGNs,  galactic black hole  and NS  systems, that
broad iron  fluorescence lines are  produced by reflection  of photons
from a  hot corona on  the surface of  an accretion disk  that extends
very close to the compact object.

\section{Observation, data reduction and spectral selection criteria} 

GX  340+0  was observed  with  XMM-Newton  from  2007 September  2  at
13:18:00 to September 3 at 02:32:05 UTC, for a total observing time of
47640 s (Obs.ID  0505950101). However, because of the  high count rate
reached  in the  EPIC-PN  CCDs  during most  of  the observation,  the
effective  exposure  was  reduced  by  data losses  due  to  telemetry
saturation. The total  good time intervals, thus, resulted  in 40562 s
effective exposure. In this work  we use data collected by the EPIC-PN
instrument  in fast  timing mode;  the source  region  (the background
region) has been selected using CCD rows RAWX = 28-48 and RAWY = 0-200
(RAWX  = 2-8  and RAWY  = 0-200).  No background  flaring  was present
during the observation.  We selected only events with PATTERN $\leq$ 4
(singles and doubles) and FLAG =  0 and restricted our analysis to the
energy range 2.2–11.8 keV.  We excluded data from PN below 2.2 keV and
rely on  the Reflection Grating  Spectrometers (RGS1 and RGS2,  in the
0.4--2.0 keV  energy band)  for the soft energy band, because  PN data
present  strong  systematic  residuals  of instrumental  origin  which
affect  the  overall  continuum  determination.   PN  data  have  been
progressively  rebinned  to  avoid  data oversampling;  RGS  data  are
rebinned  in order  to have  at least  25 counts  per  energy channel.
Source, background spectra and  response matrices were extracted using
the XMM-Newton Science Analysis Software (SAS, ver.  7.1.0).  Spectral
analysis was  performed using  Xspec ver.  12.4.0.   Temporal analysis
was performed using the FTOOLS ver. 6.4.1.

\begin{figure}
\centering
  \includegraphics[angle=-90, width=12cm]{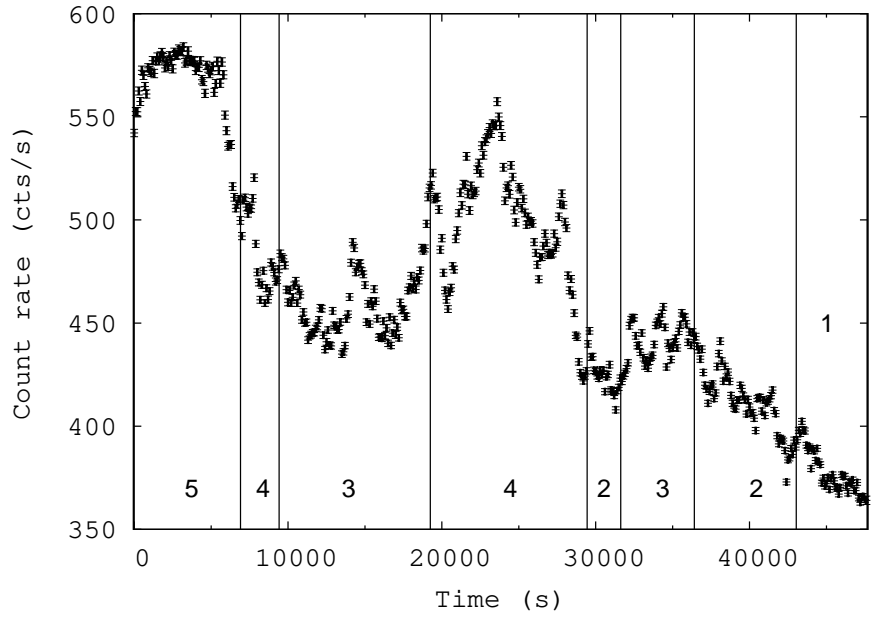}
  \caption{Light curve of GX 340+0 from the EPIC-PN, starting from MJD 54345.554. 
Bin time 100 s. Vertical lines and numbers inside
the boxes indicate the time selection for the corresponding energy spectra and the PSD shown in Figure 2.}
\label{fig1}
\end{figure}

In  Figure \ref{fig1} we  show the  light curve  obtained by  the EPIC-PN
instrument  in  the  0.5--10  keV  energy  range.   The  source  shows
significant  X-ray variability,  with a  40\% decrease  in  count rate
between the beginning and end of the observation.

\begin{figure}
\centering
  \includegraphics[width=12cm]{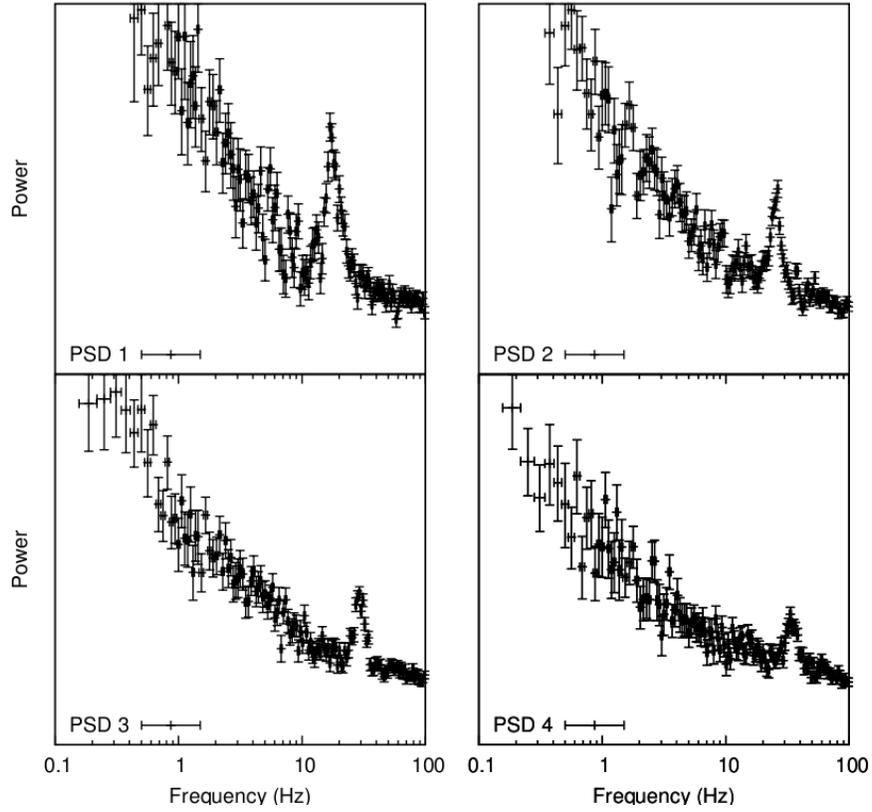}
  \caption{Lehay normalized power density spectra for four time-selected intervals, corresponding
 to the time selections 1-4 used in the spectral analysis. Peak frequencies of the HBO are reported in Table 1.}
\label{fig2}
\end{figure}

Besides  the  spectral analysis  we  also  performed  a study  of  the
temporal variability.   We extracted power spectral densities (PSD) from
the EPIC-PN data in the frequency range 1/16 Hz to 512 Hz. We obtained
a single PSD after averaging every  128 PSD, thus having a total of 20
PSD for the entire observation.   We fitted the averaged PSD using two
zero-centered Lorentzians, to account  for the broadband low frequency
noise, a constant term to fit  the white noise and a Lorentzian to fit
a broad quasi-periodic oscillation.  We identify this broad QPO as the
HBO   \citep{jonker00};   this    feature,   as   clearly   shown   by
\citet{jonker00} is a good tracer of the overall temporal variability,
which in turn, can be used  to track the accretion state of the source
in the X-ray Color-Color Diagram.  With the exception of the first two
PSD, when the source was at  the highest count rate, we clearly detect
the HBO in  all the remaining PSD.  The HBO centroids  span a range of
frequencies  from 17  to 37  Hz,  and we  noted this  frequency to  be
correlated  with the  source count  rate.   We used  these results  to
derive  the   good  time  intervals   from  which  we   extracted  the
corresponding energy spectra.  After  some trials, we decided to slice
the observation into five segments; this choice allows us to trail the
spectral evolution  of the source,  still obtaining a  high statistics
spectrum  to constrain all  the spectral  parameters.  We  denote from
Spectrum 1  to Spectrum 5 the  PN and RGS energy  spectra extracted in
the time intervals when the HBO frequency was in the 17--22 Hz, 22--27
Hz,  27--32  Hz,  32--37  Hz,  $>$  37 Hz  (or  not  detected)  range,
respectively  (see  the  time   selected  intervals  shown  in  Figure
\ref{fig1}).  In  Figure \ref{fig2}, we  show the time  averaged power
spectra for  the time  intervals 1  to 4: the  HBO evolved  during the
observation,  moving from  lower  frequencies, small  widths and  high
spectral power  (Spectrum 1) to higher frequencies,  larger widths and
less spectral  power (Spectrum 4).  The HBO is probably  undetected in
Spectrum  5 because of  the lower  statistics, and  because it  is too
broad to be resolved.

Given the  high count  rate of the  source, we  carefully investigated
pile-up  related effects  by extracting  our spectra  from rectangular
regions with and  without the inclusion of the  brightest central rows
\footnote{Strictly  following  the  SAS  thread  as  described  in  in
  http://heasarc.nasa.gov/docs/xmm/sas/USG/node63.html}.        Through
several different trials, and with the help of the $epatplot$ tool, we
concluded  that spectra  1  to  4 are  not  significantly affected  by
pile-up,  and, as far  as we  are mostly  concerned with  the spectral
shape of  the broad  iron Fe K$\alpha$  line, the  spectral parameters
are,  within the  relative uncertainties,  self-consistent  for region
masks with  and without the  inclusion of the central  brightest rows.
On the other hand, the  time-selected region 5 is affected by pile-up;
we extracted, therefore, the  associated energy spectrum excluding the
central 2 brightest pixel rows,  in order to avoid significant pile-up
dependent spectral distortions.

\section{Spectral Analysis}

We  analyzed  the  time  selected spectra  independently,  adopting  a
continuum   model   consisting   of   soft   thermal   disk   emission
(\texttt{diskbb}  component,  \citealt{makishima86})  and  a  thermal,
harder, blackbody emission (component \texttt{blackbody}).  We used
for  the  interstellar   absorption  the  \texttt{vphabs}  model  with
crss-sections   of   \citet{verner96}    and   table   abundances   of
\citet{wilms00}.  We  found significant residuals at  the neutral iron
edge (7.11 keV); leaving the iron abundance free to vary resulted in a
significant  improvement   in  the  $\chi^2$  and   in  the  residuals
scattering;  iron  is  under-abundant  with  respect  to  the  assumed
\cite{wilms00}  solar value (see  also \cite{ueda05}).

Although broadband spectra of Z-sources reveal that the hard component
is  better  described  by  thermal  Comptonization  models  \citep[see
e.g.][]{lavagetto04},  XMM data  do  not allow  to discriminate  among
different  models of  X-ray continuum,  or to  reliably  constrain the
parameters of  the Comptonization model.  We tested,  however, using a
variety of different spectral models, that the choice of the continuum
alter, only marginally the overall  $\chi^2$ value of our fits and the
determination  of the  local features  parameters, which  is  the main
objective of the present Letter.

Superimposed to  the continuum emission,  a few discrete  features are
always present: a broad emission feature in the Fe K$\alpha$ region, a
weaker  but  also  broad  emission  line  at $\sim$  3.9  keV  and  an
absorption edge, whose energy threshold is in the 8.7--9.0 keV range.

We preliminarly investigated  the shape of the broad  iron line, using
the total time averaged PN spectrum (energy range 3-12 keV), excluding
only the time interval 5  (see Figure 1) because pile-up affected this
part of  the observation.   Using a Gaussian  profile to fit  the line
emission,  resulted   in  a   line  position  of   6.76$\pm$0.02  keV,
0.24$\pm$0.02  keV  width  and (2.0$\pm$0.2)$\times$10$^{-3}$  photons
cm$^{-2}$  s$^{-1}$  normalization  value.  Using  a  relativistically
smeared disk-reflected  profile (modelled using  the \texttt{diskline}
profile of \citealt{fabian89}), resulted in a rest frame energy of the
line at 6.69$\pm$0.02 keV, an inner radius R$_{in}$=13$\pm$3 R$_g$, an
inclination    angle   34.6$\pm$1.3    deg   and    emissivity   index
2.50$^{+0.09}_{-0.13}$.  The R$_{ext}$ has a best fitting value $\sim$
10$^{4}$ R$_g$ but it is  rather unconstrained with no upper limit and
a 90\%  c.l. lower limit at  3000 R$_g$.  The  $\Delta \chi^2$ between
these  two profiles  is 25  (+3  dof in  the \texttt{diskline}  model)
indicating  the relativistic  profile as  statistically  more favorite
than the symmetric Gaussian profile.

We  assumed, thereafter,  that the  3.9  keV emission  line, which  we
identify with the Ca XIX Ly$\alpha$ transition is also produced in the
disk reflecting plasma, and  we fit it using another \texttt{diskline}
component, with  R$_{in}$ \footnote{Inner and outer disk  radii in the
  \texttt{diskline}  model  are expressed  in  units of  gravitational
  radii (R$_{g}$), which for a  1.4 solar masses NS corresponds to 2.1
  km.}  , R$_{ext}$, emissivity index  and inclination bound to the Fe
K$\alpha$ \texttt{diskline} component.

We  turned,  then, to  the  analysis  of  the time  selected  spectra.
Because of  the high statistics it  is possible to  well constrain all
the \texttt{diskline} spectral parameters, except the outer radius for
which we derived the largest uncertainties. We fixed this parameter to
the reference  value of  10$^4$ R$_g$, after  having tested  that this
value lies  extremely close ($\Delta  \chi^2 \leq$ 1) to  the $\chi^2$
minimum for  each spectrum.  We find  that the line  shape, within the
statistical uncertainties,  does not significantly  change during most
of the observation, despite a significant change of the continuum flux
and  continuum  parameters  (see  Table\ref{tab1})  with  best fitting
values all consistent with the total time averaged values.

\begin{figure}
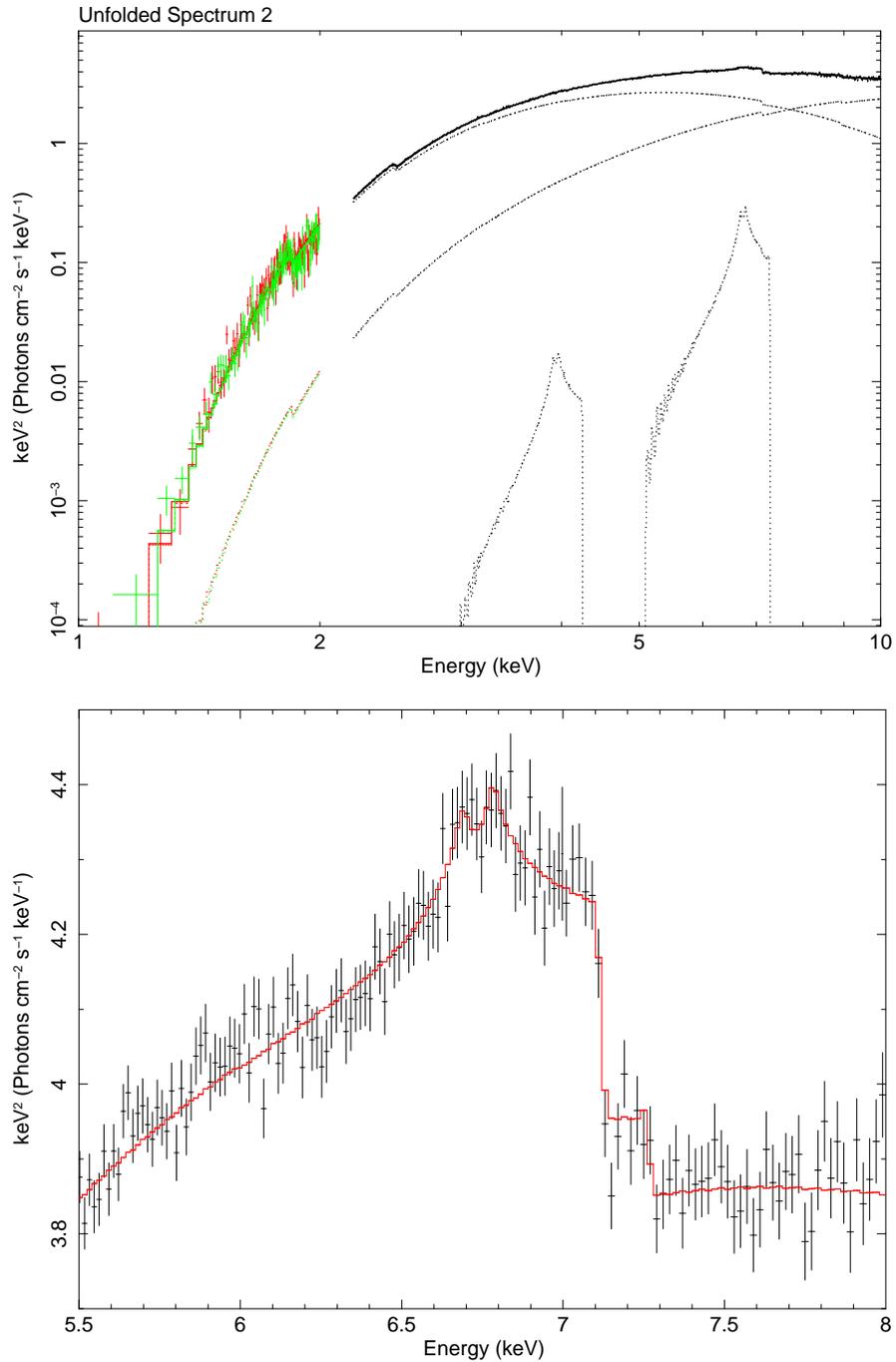

\centering
\begin{tabular}{c}
\includegraphics[angle=-90, width=12cm]{f3a.eps}\\
\includegraphics[angle=-90, width=12cm]{f3b.eps}\\
\end{tabular}
\caption{Top panel: unfolded spectrum (in the f(E)*E$^2$ representation, together with the best-fit model
in the 1.0-10.0 keV band for the representative Spectrum 2. 
Bottom panel: snapshot of the top panel figure in the 5.5--7.5 restricted energy band.}
\label{fig3}
\end{figure}

For Spectra 4 and 5, corresponding to the highest luminosities reached
by  the  source  during   the  observation,  the  temperature  of  the
\texttt{blackbody} component  was rather unconstrained by  the fit, so
that we kept it frozen in the fits to a reference value of 3.5 keV, in
order to avoid unphysical higher values due to the lack of high energy
spectral  coverage. In  Table 1  we comprehensively  show  the fitting
results for the five  time selected spectra.  We representatively show
in Figure  \ref{fig3} the unfolded spectrum, with  the contribution of
the additive components and a snapshot of the iron region, of Spectrum
2.

\section{Discussion and Conclusions}

In this Letter we report on the first unambiguous determination of the
broad disk  origin of the  iron line in  the Z-source GX  340+0. Broad
asymmetric iron lines  present in the energy range  6.4--6.97 keV have
been  discovered first  in AGNs,  and  later, in  galactic black  hole
candidates,    and   very    recently    also   in    a   NS    system
\citep{disalvo05,bhattacharyya07,cackett08}.  The  common scenario for
the  origin  of  such  lines   is  disk  reflection,  where  the  line
deformation primarily  arises from the high  Keplerian velocity fields
of the  disk reflecting matter  and from general  relativistic effects
produced by the intense gravitational  well of the compact object. Our
analysis  reveals,  for  the first  time  in  a  NS accreting  at  the
Eddington limit that  the iron line shape in  GX 340+0 is unambigously
described by a disk reflection profile, finding also evidence of other
reflection  signatures (an  absorption edge  at $\sim$  8.8 keV  and a
broad Ca XIX emission line).

During this XMM-Newton observation  we can closely follow the spectral
evolution of  GX 340+0 through all  of its horizontal  branch.  At its
leftmost  point, corresponding  to  the  lowest count  rate  and to  a
frequency  of  the  HBO  of   $\sim$  17  Hz,  we  derive  the  lowest
temperatures both for the disk  (1.60 keV) and the hard emission (2.55
keV); as the source moves towards  the hard apex, there is a continuum
rising of the temperatures and  of the X-ray flux; assuming a distance
of 11$\pm$3  kpc and an inclination  of the disk as  inferred from the
diskline profile,  we derive  a change in  the source  luminosity from
(1.67$\pm$0.1) $\times$ 10$^{38}$ erg/s (Spectrum 1) to (2.57$\pm$0.2)
$\times$  10$^{38}$  erg/s (Spectrum  5).  The  inner  disk radius  as
derived from  the \texttt{diskbb} normalization (see Table  1) gives a
value in  the 8--11 km range;  considering that the  inner disk radius
can be  underestimated by a  factor of two \citep{merloni00},  we find
that  these values are  in agreement  with the  ones derived  from the
\texttt{diskline} line profile.

Despite  this remarkable continuum  change, the  profile of  the broad
iron diskline does not sensibly  change.  The rest frame energy of the
\texttt{diskline} component present in  the iron region indicates that
the line  is most probably  produced by He-like  iron ions at  a short
distance  from  the  compact   source.  In  Spectrum  5,  because  the
inclination angle and the emissivity  of the line were bad constrained
by  the fit,  we  kept them  frozen to  the  best values  of the  time
averaged  spectrum; we  noted, therefore,  a small  shift of  the rest
frame  line  energy towards  higher  energies,  possibly indicating  a
greater contribution from the H-like ions.

The equivalent width of the  line decreases from Spectrum 1 ($\sim$ 60
eV) to  Spectrum 5 ($\sim$ 30  eV), indicating that the  line does not
respond to the soft X-ray  flux, which, on the contrary increases, but
presumably to the  hard X-ray emission, above 12  keV, which falls out
from the  observed spectral coverage  of the EPIC/PN  instrument.  The
inclination angle of  the disk and the emissivity  index are very well
constrained by  our fits,  with relative uncertainties  of only  a few
percent.   Contrary  to  what  found  in  galactic  black  holes,  the
emissivity  index  is  not  very  steep, indicating  a  more  extended
illuminating corona  above the disk; we  are not able  to identify any
significant  change  in  the   value  of  this  parameter  during  the
observation,  despite a  large  increase in  the  temperatures of  the
continuum components and their related fluxes.

Recently  \citet{laurent07} proposed an  alternative scenario  for the
formation  of broad  iron lines  in LMXBs,  where extensive  red wings
could be formed by recoil of line photons in an optically thick medium
expanding,  or  converging,   at  relativistic  velocities.   Since  a
spectral line  model, adapted to  this scenario, is not  yet available
for fitting  X-ray data, we are  not able to test  this scenario using
our data. However,  as pointed in \citet{pandel08} for  the case of 4U
1636-536, we note that the lack of a narrow core and the presence of a
blue wing in the iron line  profile of GX 340+0 is a strong indication
of the disk reflection origin.

Most  theoretical models on  the QPO  generation in  NS LMXB  agree to
identify the shortest dynamical timescales, i.e. the highest frequency
QPO at  $\sim$ 1 kHz,  with the frequency  of the Keplerian  motion of
matter at the inner rim of  an accretion disk.  From the HBO frequency
and the  relation known to  exist between this  and the upper  kHz QPO
frequency \citep{jonker00},  we can infer the  upper kHz QPO  to be in
the 550--750 Hz frequency range. For a typical 1.4 M$_{\odot}$ NS this
frequency  range corresponds  to  an inner  disk  radius of  9.8--12.1
R$_g$.   The  \texttt{diskline}  inferred  inner radii  for  the  time
selected spectra are all fully  consistent within this range, and give
support to the identification of  highest QPO frequency with the inner
disk  Keplerian frequency.  Our fit  results seem  also to  indicate a
decrease in  the value of  the inner disk  radius as the  source moves
towards higher luminosities.  The  errors associated to this parameter
are, however, still too large to claim a strict correlation and longer
exposures are needed to assest the significance of such a relation.

The spectra show, together with  the broad Fe K$\alpha$ emission line,
other reflection signatures, like a  broad Ca XIX resonance line and a
high-energy absorbtion  edge, compatible with the Fe  XXV K-edge (8.83
keV energy  in the laboratory frame).  The  extraordinary good quality
of the  EPIC-PN spectrum of this  source will allow to  test even more
recent  theoretical  line  profiles  and  fully  self-consistent  disk
reflection models.  We aim at expanding this analysis in a forthcoming
paper.

\acknowledgements Authors express their  gratitude to the EPIC/PN team
and,  especially, to  Matteo Guainazzi  for  the kind  assistance
offered us in  the process of data extraction and  reduction of the PN
data. Authors would like to thank the anonymous referee for the carefull
reading of the manuscript and the many suggestions.
Authors also thank Giuseppe Lavagetto
for useful discussions.

\bibliographystyle{apj}
\bibliography{refs}

\begin{thebibliography}{18}
\expandafter\ifx\csname natexlab\endcsname\relax\def\natexlab#1{#1}\fi

\bibitem[{{Berendsen} {et~al.}(2000){Berendsen}, {Fender}, {Kuulkers}, {Heise},
  \& {van der Klis}}]{berendsen00}
{Berendsen}, S.~G.~H., {Fender}, R., {Kuulkers}, E., {Heise}, J., \& {van der
  Klis}, M. 2000, \mnras, 318, 599

\bibitem[{{Bhattacharyya} \& {Strohmayer}(2007)}]{bhattacharyya07}
{Bhattacharyya}, S., \& {Strohmayer}, T.~E. 2007, \apjl, 664, L103

\bibitem[{{Cackett} {et~al.}(2008){Cackett}, {Miller}, {Bhattacharyya},
  {Grindlay}, {Homan}, {van der Klis}, {Miller}, {Strohmayer}, \&
  {Wijnands}}]{cackett08}
{Cackett}, E.~M., {Miller}, J.~M., {Bhattacharyya}, S., {Grindlay}, J.~E.,
  {Homan}, J., {van der Klis}, M., {Miller}, M.~C., {Strohmayer}, T.~E., \&
  {Wijnands}, R. 2008, \apj, 674, 415

\bibitem[{{Di Salvo} {et~al.}(2005){Di Salvo}, {Iaria}, {M{\'e}ndez},
  {Burderi}, {Lavagetto}, {Robba}, {Stella}, \& {van der Klis}}]{disalvo05}
{Di Salvo}, T., {Iaria}, R., {M{\'e}ndez}, M., {Burderi}, L., {Lavagetto}, G.,
  {Robba}, N.~R., {Stella}, L., \& {van der Klis}, M. 2005, \apjl, 623, L121

\bibitem[{{Fabian} {et~al.}(1989){Fabian}, {Rees}, {Stella}, \&
  {White}}]{fabian89}
{Fabian}, A.~C., {Rees}, M.~J., {Stella}, L., \& {White}, N.~E. 1989, \mnras,
  238, 729

\bibitem[{{Hasinger} \& {van der Klis}(1989)}]{hasinger89}
{Hasinger}, G., \& {van der Klis}, M. 1989, \aap, 225, 79

\bibitem[{{Jonker} {et~al.}(2000){Jonker}, {van der Klis}, {Wijnands}, {Homan},
  {van Paradijs}, {M{\'e}ndez}, {Ford}, {Kuulkers}, \& {Lamb}}]{jonker00}
{Jonker}, P.~G., {van der Klis}, M., {Wijnands}, R., {Homan}, J., {van
  Paradijs}, J., {M{\'e}ndez}, M., {Ford}, E.~C., {Kuulkers}, E., \& {Lamb},
  F.~K. 2000, \apj, 537, 374

\bibitem[{{Laurent} \& {Titarchuk}(2007)}]{laurent07}
{Laurent}, P., \& {Titarchuk}, L. 2007, \apj, 656, 1056

\bibitem[{{Lavagetto} {et~al.}(2004){Lavagetto}, {Iaria}, {di Salvo},
  {Burderi}, {Robba}, {Frontera}, \& {Stella}}]{lavagetto04}
{Lavagetto}, G., {Iaria}, R., {di Salvo}, T., {Burderi}, L., {Robba}, N.~R.,
  {Frontera}, F., \& {Stella}, L. 2004, Nuclear Physics B Proceedings
  Supplements, 132, 616

\bibitem[{{Makishima} {et~al.}(1986){Makishima}, {Maejima}, {Mitsuda}, {Bradt},
  {Remillard}, {Tuohy}, {Hoshi}, \& {Nakagawa}}]{makishima86}
{Makishima}, K., {Maejima}, Y., {Mitsuda}, K., {Bradt}, H.~V., {Remillard},
  R.~A., {Tuohy}, I.~R., {Hoshi}, R., \& {Nakagawa}, M. 1986, \apj, 308, 635

\bibitem[{{Merloni} {et~al.}(2000){Merloni}, {Fabian}, \& {Ross}}]{merloni00}
{Merloni}, A., {Fabian}, A.~C., \& {Ross}, R.~R. 2000, \mnras, 313, 193

\bibitem[{{Oosterbroek} {et~al.}(1994){Oosterbroek}, {Lewin}, {van Paradijs},
  {van der Klis}, {Penninx}, \& {Dotani}}]{oosterbroek94}
{Oosterbroek}, T., {Lewin}, W.~H.~G., {van Paradijs}, J., {van der Klis}, M.,
  {Penninx}, W., \& {Dotani}, T. 1994, \aap, 281, 803

\bibitem[{{Pandel} {et~al.}(2008){Pandel}, {Kaaret}, \& {Corbel}}]{pandel08}
{Pandel}, D., {Kaaret}, P., \& {Corbel}, S. 2008, ArXiv e-prints

\bibitem[{{Penninx} {et~al.}(1993){Penninx}, {Zwarthoed}, {van Paradijs}, {van
  der Klis}, {Lewin}, \& {Dotani}}]{penninx93}
{Penninx}, W., {Zwarthoed}, G.~A.~A., {van Paradijs}, J., {van der Klis}, M.,
  {Lewin}, W.~H.~G., \& {Dotani}, T. 1993, \aap, 267, 92

\bibitem[{{Schulz} \& {Wijers}(1993)}]{schulz93}
{Schulz}, N.~S., \& {Wijers}, R.~A.~M.~J. 1993, \aap, 273, 123

\bibitem[{{Ueda} {et~al.}(2005){Ueda}, {Mitsuda}, {Murakami}, \&
  {Matsushita}}]{ueda05}
{Ueda}, Y., {Mitsuda}, K., {Murakami}, H., \& {Matsushita}, K. 2005, \apj, 620,
  274

\bibitem[{{Verner} {et~al.}(1996){Verner}, {Ferland}, {Korista}, \&
  {Yakovlev}}]{verner96}
{Verner}, D.~A., {Ferland}, G.~J., {Korista}, K.~T., \& {Yakovlev}, D.~G. 1996,
  \apj, 465, 487

\bibitem[{{Wilms} {et~al.}(2000){Wilms}, {Allen}, \& {McCray}}]{wilms00}
{Wilms}, J., {Allen}, A., \& {McCray}, R. 2000, \apj, 542, 914

\end{thebibliography}
\clearpage

\clearpage

\clearpage

\begin{deluxetable}{lrrrrr}
\tabletypesize{\small}
\tablewidth{0pt}
\tablecolumns{7}
\tablecaption{Spectral fitting results}
\tablehead{
\colhead{} & 
\colhead{Spectrum 1} & 
\colhead{Spectrum 2}&
\colhead{Spectrum 3}& 
\colhead{Spectrum 4}& 
\colhead{Spectrum 5}}
\startdata 
HBO freq. (Hz)                      &  17.7$\pm$0.2  & 25.7$\pm$0.2   & 29.4 $\pm$0.2   & 34.6$\pm$0.4  &   37-40\tablenotemark{A} \\
0.5-2.0 keV Flux\tablenotemark{B}   & 2.69           & 2.99           & 3.22            & 3.55          & 4.17  \\
2.0-10.0 keV Flux\tablenotemark{B}  & 9.38           & 10.45          & 11.51           & 12.77         & 14.82  \\

$N_{\rm H}$\tablenotemark{C}         & 9.42$_{-0.21}^{+0.26}$  & 9.84$ \pm 0.20$          & 9.89$\pm 0.19$                                             & 10.23$\pm 0.14$     & 10.51$\pm 0.27$  \\
Iron Abundance             & 0.84$_{-0.2}^{+0.13}$   & 0.75 $\pm 0.11$          & 0.64 $\pm$ 0.07                                             & 0.72 $\pm$ 0.08     & 0.68 $\pm$0.13                          \\ 

$kT_{disk}$ (keV)                    & 1.60$_{-0.11}^{+0.23}$  & 1.85$\pm$0.09    & 1.88$\pm$0.08                                                  & 2.06$_{-0.02}^{+0.03}$ & 2.07$\pm$0.3 \\
R$_{DBB}$\tablenotemark{D} (km)      & 9.7$_{-1.7}^{+1.3}$     & 7.8$ \pm 0.6$           & 7.6$ \pm 0.4 $                                                  & 8.0$\pm$0.4 & 10.5$\pm 0.4$ \\

$kT_{bb}$ (keV)                      & 2.55$_{-0.12}^{+0.22}$  & 2.89$_{-0.20}^{+0.23}$    & 2.98$\pm$0.20                                                 & $>$3.5 & $>$3.5  \\

Fe E$_{line}$ (keV)                  & 6.70$\pm 0.03$         & 6.72$\pm 0.03$           & 6.67$\pm 0.03$                                                &  6.72$\pm$0.03 &  6.78$_{-0.05}^{+0.06}$ \\
R$_{in}$  (R$_g$)                    & 16$_{-8}^{+20}$         & 13$_{-7}^{+5}$            & 14$\pm$2                                                     & 13$_{-7}^{+6}$  & 7$_{-1}^{+18}$  \\
R$_{ext}$\tablenotemark{E}(R$_g$)    & 10$^4$                 &  10$^4$                   &  10$^4$                                                      &  10$^4$   & 10$^4$ \\
Incli.  (deg)                       &  40$_{-5}^{+10}$        & 33$_{-2.0}^{+2.5}$         & 36.2$_{-1.3}^{+0.9}$                                          & 35.7$\pm$1.7   &35\tablenotemark{E}  \\
Betor \tablenotemark{F}             & 2.38$_{-0.10}^{+0.14}$  & 2.46$_{-0.07}^{+0.11}$      & 2.52$\pm$0.07                                                & 2.47$\pm$0.07 & 2.50\tablenotemark{E} \\
Fe Norm.\tablenotemark{G}           & 5.5$_{-0.8}^{+0.6}$     & 5.1$\pm$1.0              & 5.7$\pm$0.5                                                  & 4.0$\pm$0.6 & 5.6$_{-1.2}^{+3}$\\
Line Equivalent Width (eV)          &  60                    &   43                       &  43                                                            &    33         &      33          \\
Ca E$_{line}$ (keV)                 & 3.87$\pm$0.07   & 3.93$_{-0.10}^{+0.17}$      & 3.94$\pm$0.05                                                & 3.94$\pm$0.06 & 3.92$\pm$0.07 \\
Ca Norm.\tablenotemark{G}           & 0.55$\pm$0.06  & 0.63$\pm$0.1         & 1.5$\pm$0.3                                                 & 1.4$_{-0.9}^{+1.9}$ & 3.7$_{-1.4}^{+2.7}$ \\
Line Equivalent Width (eV)          &    4.2                 &    2.2                      &  3.0                                                            & 4.1            &  6.0              \\
Edge E (keV)                        & 8.85$\pm 0.10$    & 8.97$\pm 0.10$                  & 8.94$\pm0.10$                                              & 8.80$_{-0.07}^{+0.10}$ &  8.91$_{-0.09}^{+0.22}$ \\
Edge $\tau$ ($\times 10^{-2}$)      & 4$\pm 1$           & 3$\pm 1$                       & 3$\pm 1$                                                   &  3.0$_{-0.7}^{+0.6}$ & 3.7$\pm$1.6 \\

\hline

$\chi^2_{red}$ ($dof$)              & 0.974 (674)              &    1.083 (771)    &   1.244 (908)                                                        &  1.186 (866)  & 1.090 (663)  \\

\enddata  
\tablenotetext{A}{QPO not resolved in the PSD, value inferred to be in this range from \cite{jonker00}. See also Fig.2}
\tablenotetext{B}{Unabsorbed flux in units $10^{-9}$ ergs cm$^{-2}$ s$^{-1}$. Reported values have
a $\sim$ 10\% relative error.}
\tablenotetext{C}{In units of $10^{22}$ atoms cm$^{-2}$.}
\tablenotetext{D}{Inner disk radius in km, as derived from the normalization parameter $N$ of the \texttt{diskbb} component: $R_{DBB}= 
D \sqrt{N/cos(\theta)}$; for the calculation we assume a distance ($D$) of 11 kpc and an inclination angle ($\theta$) of 35 deg.}
\tablenotetext{E}{Frozen parameter during the fitting procedure.}
\tablenotetext{F}{Index of the emissivity power law, that scales as r$^{-(Betor)}$.}
\tablenotetext{G}{Normalization values of the \texttt{diskline} component in units of $10^{-3}$ photons cm$^{-2}$ s$^{-1}$.}
\tablecomments{Best-fitting values and associated errors for spectra 1-5. Errors quoted at $\Delta \chi^2=2.7$.}
\label{tab1}
\end{deluxetable}

\end{document}